\RequirePackage{fix-cm}
\documentclass[twocolumn]{svjour3}          
\smartqed  
\usepackage[utf8]{inputenc}
\usepackage{graphicx}
\usepackage[dvipsnames]{xcolor}
\usepackage{hyperref}
\hypersetup{colorlinks=true, linkcolor=blue, citecolor=ForestGreen, urlcolor=DarkOrchid}

\usepackage{xurl}
\usepackage{amsmath}
\usepackage{amssymb}
\usepackage{physics}
\usepackage{mathtools}
\usepackage{pgfmath}
\usepackage{tikz}
\usepackage{adjustbox}
\usetikzlibrary{calc, 
   arrows,
   shapes,
   graphs,
 fadings, 
 positioning,
 decorations.markings, 
 intersections, 
 backgrounds,
 fit,
   circuits.logic.US,
   tikzmark,
   zx-calculus,
   external
}

\begin{document}

\title{QUASIM -- Quantum Computing Enhanced Service Ecosystem for Simulation in Manufacturing 
}

\titlerunning{QUASIM -- Quantum Computing Enhanced Service Ecosystem for Simulation in Manufacturing}        

\author{Wolfgang Maass$^{1,2}$         \and
Ankit Agrawal$^1$         \and
Alessandro Ciani$^3$         \and
Sven Danz$^{3,4}$         \and
Alejandro Delgadillo$^{5}$         \and
Philipp Ganser$^{6}$         \and
Pascal Kienast$^{6}$         \and
Marco Kulig$^7$         \and
Valentina König$^{5}$         \and
Nil Rodellas-Gr\`acia$^{3,4}$   \and
Rivan Rughubar$^{3,4}$         \and
Stefan Schröder$^{6}$         \and
Marc Stautner$^{5,8}$         \and
Hannah Stein$^{1,2}$         \and
Tobias Stollenwerk$^3$         \and
Daniel Zeuch$^3$         \and
Frank K. Wilhelm$^{3,4}$
}

\authorrunning{Wolfgang Maass et al.} 

\institute{Corresponding author: Wolfgang Maass \at
              \email{wolfgang.maass@dfki.de}
           \\
           $^1$German Research Center for Artificial Intelligence, Stuhlsatzenhausweg 3, 66123 Saarbrücken, Germany 
           \\
           $^2$Saarland University, Campus A5 4, 66123 Saarbrücken, Germany 
           \\
           $^3$Institute for Quantum Computing Analytics (PGI-12), Forschungszentrum Jülich, D-52425 Jülich, Germany
           \\
           $^4$Theoretical Physics, Saarland University, 66123 Saarbrücken, Germany
           \\
           $^5$ModuleWorks GmbH, Henricistraße 50-52, 52072 Aachen, Germany
           \\
           $^6$Fraunhofer Institute for Production Technology IPT, Steinbachstraße 17, 52074 Aachen, Germany 
           \\
           $^7$TRUMPF Werkzeugmaschinen SE + Co. KG, Johann-Maus-Straße 2, 71254 Ditzingen, Germany
             \\
           $^8$Hochschule Ruhr West, Duisburger Straße 100, 45479 Mülheim an der Ruhr, Germany
}

\date{Received: date / Accepted: date}

\maketitle
\begin{abstract}
Quantum computing (QC) and machine learning (ML), taken individually or combined into quantum-assisted ML (QML), are ascending computing paradigms whose calculations come with huge potential for speedup, increase in precision, and resource reductions. Likely improvements for numerical simulations in engineering imply the possibility of a strong economic impact on the manufacturing industry. In this project report, we propose a framework for a quantum computing-enhanced service ecosystem for simulation in manufacturing, consisting of various layers ranging from hardware to algorithms to service and organizational layers. In addition, we give insight into the current state of the art of applications research based on QC and QML, both from a scientific and an industrial point of view. We further analyze two high-value use cases with the aim of a quantitative evaluation of these new computing paradigms for industrially relevant settings.
\end{abstract}

\section{Introduction}
\label{sec:intro}

Quantum computing (QC), quantum sensing, and quantum communication are emergent technologies that are expected to have a disruptive effect on large parts of science, technology, and the economy at large \cite{Acin18}.  Of these three quantum technologies, quantum computing is thought of as having the greatest potential for economic impact \cite{acatech20lang_eng} as it promises speedup resulting in a change of time complexity for certain hard computational tasks \cite{nielsenChuang}. Quantum computing technologies are evaluated in various domains, such as finance \cite{egger2020quantum}, production \cite{bayerstadler2021industry}, chemical modeling \cite{McArdle20}, and drug design \cite{blunt2022perspective}. Some algorithms feature provable speedup \cite{nielsenChuang}, while many others are being analyzed by simulation on classical computers or early QCs \cite{cross2018ibm,bergholm2018pennylane}. 
Classical simulations of QC algorithms have, in general, exponential resource requirements that grow with the number of qubits and operations. While quantum algorithms of optimization tasks as targeted by operations research (e.g., scheduling and logistics) is a busy field of research, little attention has been paid to tasks in manufacturing (e.g., finite element methods (FEM)) \cite{montanaroPallister2016}.

The success of artificial intelligence (AI) technologies, in particular machine learning (ML), in the areas of natural language processing \cite{vaswani2017attention} and visual computing \cite{krizhevsky2012imagenet} is based on processing excessive amounts of data for training ML models. Large infrastructures are required for parallel execution of simple operations followed by optimization of model weights according to objective functions \cite{goodfellow2016deep}. Training large ML models currently takes several weeks. Hence, updating such models becomes a resource challenge. Quantum computing technologies could become a means for accelerating ML processes with the goal of real-time training of ML models with limited resource requirements. This will provide enormous economic advantages if realized. The combination of machine learning with quantum computing, known as quantum-assisted machine learning or simply quantum machine learning (QML) \cite{biamonte2017,schuld2015}, harbors the potential of increased computing power for solving complex optimization problems \cite{biamonte2017}. With a competition of established scientific computing and ML in the classical realm, we can also ask in the QC domain which approach is more promising in the long run and in the NISQ era \cite{Preskill18}. 

In this project report,\footnote{This work is part of the research project QUASIM, funded by the Federal Ministry for Economic Affairs and Climate Action (BMWK), grant number: 01MQ22001A.\\ \url{https://www.quasim-project.de/}} we discuss the applicability of QC and various approaches for QML for the simulation of manufacturing processes. We distinguish between approaches that solely rely on quantum algorithms and hybrid approaches combining QC algorithms with classical algorithms. We describe the typical bottlenecks of current computing for application in real-world tasks such as manufacturing, and we describe several strategies pursued to improve the quality and speed of those calculations. A primary goal is to develop potential use cases for QC and QML for the manufacturing industry within the lightweight design sector. The term `simulation' refers to the use of computational models to imitate physical systems over time \cite{mourtzis2014simulation}.  The present work does not encompass `quantum simulation,' in which the manipulation of a controllable quantum system directly simulates \emph{another quantum system} of interest \cite{Acin18,Georgescu13}.  The objective is to understand, develop, and evaluate software that addresses the application of QC in distinct and explicit industry-driven use cases. For science, we are working in an area where results on quantum speedup are conditional on various requirements or are only conjectured or empirical. For applications, mapping use cases to the library of quantum algorithmic primitives is an important step toward using and evaluating this technology. 


Two use cases are used to illustrate the opportunities and challenges of the practical application of QC. The first use case relates to the simulation of manufacturing processes with the aid of FEM for the example of the production of blade integrated disks (blisks) for jet engines. This involves the simulation and optimization of milling processes while complying with quality specifications, i.e., it has to be ensured that the milling process is not compromised by resonant mode excitations. In this use case, we investigate how quantum algorithms accelerate classical simulation processes. The second use case relates to the field of laser cutting of metals with a focus on heat dispersion and thermal expansion. Here, hybrid and pure QML approaches are discussed. Both cases are attractive for QC -- both in the scientific computing and the QML context. They deal with large sets of data but a relatively small set of input and output information. Also, they are based on linear or near-linear theories.

The remainder of this paper is organized as follows.  Section \ref{sec:framework} describes a framework for a QC-enhanced ecosystem for simulation in manufacturing.  Next, a number of QC- and QML-approaches to accelerating simulations in manufacturing are discussed.  This includes the description of the current state of the art in both quantum simulation (Sec.~\ref{sec:QC4CS-SOTA}) and quantum machine learning (Sec.~\ref{sec:QML-SOTA}), in each case followed by the quantum-enhanced approach (Secs.~\ref{sec:QC4CS-GeneralApproach} and \ref{sec:QML-GeneralApproach}, respectively).  To exemplify these concepts, Sec.~\ref{sec:QC4CS-SpecificApproach} describes the application of QC to a milling use case, while Sec.~\ref{sec:QML-SpecificApproach} describes the application of QML to the simulation of heat dissipation during laser cutting.

\section{Framework for QC-enhanced service ecosystem for simulation in manufacturing}
\label{sec:framework}

The goal of the project ``QC-Enhanced Service Ecosystem for Simulation in Manufacturing" (QUASIM) is to develop and test algorithms and technologies of quantum computing for critical simulation challenges in manufacturing. This involves integrating QC into Industry 4.0 frameworks as ``Quantum-as-a-Service" (QaaS) and facilitating knowledge transfer for production-oriented simulation based on QC. 

\begin{figure*}[t]
    \centering
    \includegraphics[width=0.8\linewidth]{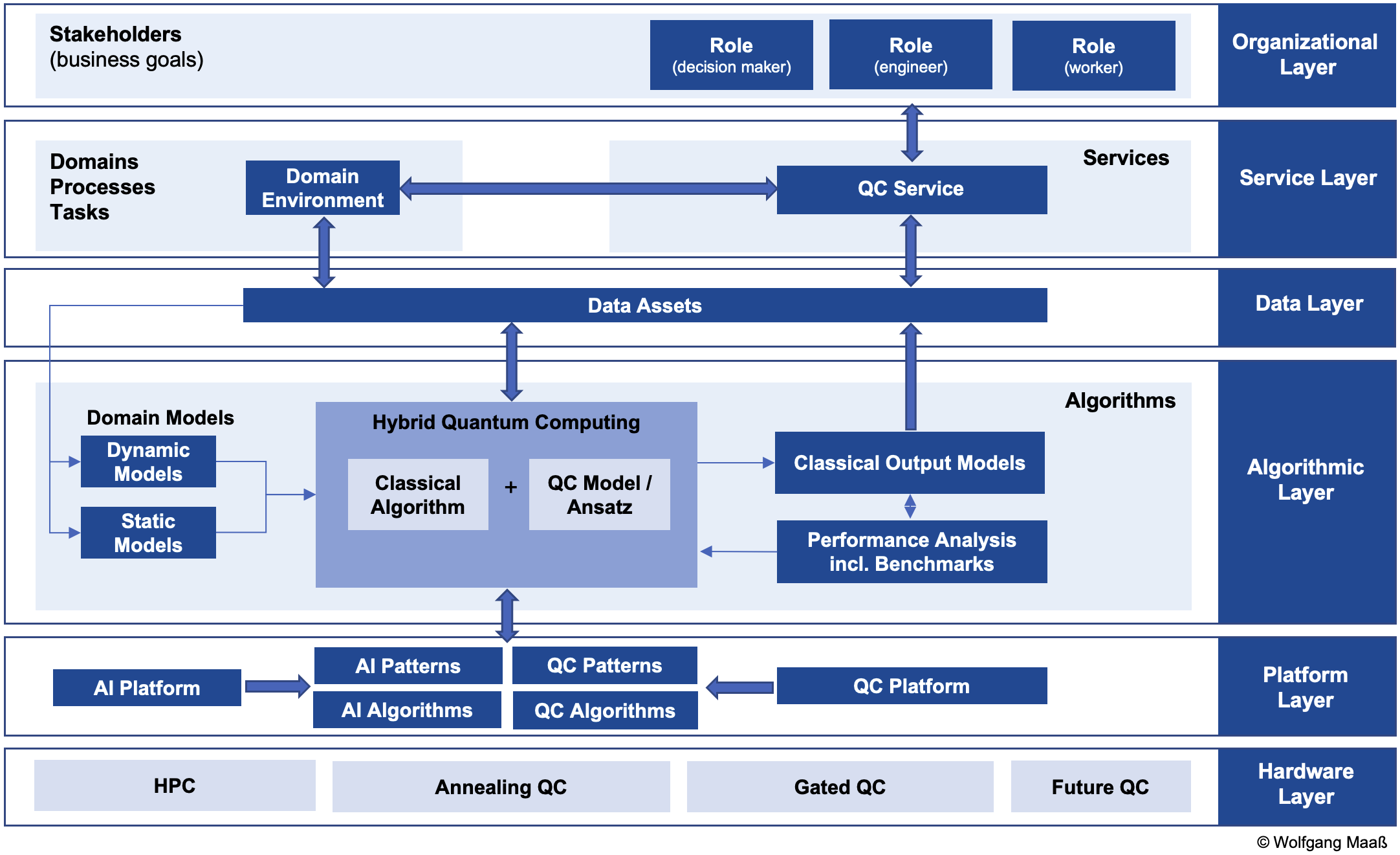}
    \caption{Envisioned QC-enhanced service ecosystem for simulation in manufacturing.}
    \label{fig:QC_SE_Framework}
\end{figure*}

Figure \ref{fig:QC_SE_Framework} shows our proposed framework for a QC-enhanced service ecosystem for simulations in manufacturing, composed of multiple layers: hardware, platform, algorithmic, data, service, and organizational. The hardware layer forms the basis for the QC-enhanced ecosystem, containing the hardware necessary for realizing QC and QML applications. The platform layer then connects AI and QC platforms via AI and QC patterns as well as algorithms. In the algorithmic layer, algorithms for both pure and hybrid QC (the combination of QC with classical computing) are developed based on AI, QC and domain models. The last can be either static or dynamic and develop from domain-specific processes and tasks, which are fed into the models via representations of data assets. This layer also hosts algorithmic benchmarks. Finally, the algorithmic results are transferred to the data layer using classical output models. 


The service layer provides QC services that are developed based on algorithms and data assets. Quantum computing services shall be made available for domain environments, e.g. tool and mold making, semiconductor industry, or engine construction, and support the processes and tasks therein. The services can be used by and provide value for different stakeholders with disparate business goals (cf. organizational layer). The underlying business goals of the stakeholders such as decision makers, engineers, or workers are included in the development and refinement of QC services to create the most value for them. QC requirements engineering methods \cite{Stein2024TowardsRE} can be used to conceptualize the services and related software in a first step.

Implemented QC-based services will be integrated into the existing digital twin framework. This opens the door to QC applications in existing digital infrastructures for industrial end users and drives QC use in industrial environments. The framework builds a basis for demonstrating practical QC applications in manufacturing and to derive medium and long-term competitive advantage for manufacturing companies using QC.

The algorithms developed within the project's scope are hardware agnostic; this feature provides flexibility in selecting appropriate hardware for each use case.  Access to QC hardware can achieved via configurable API. 

The simulation services developed will be incorporated into preexisting industry 4.0 frameworks to accelerate users' adoption. One example of such a framework is dPart ®, developed at the Fraunhofer IPT for digital twins in machining. Additionally, a QaaS backend will be available as software modules that can be integrated by CAx software providers and machine tool manufacturers into their product offerings.
In the short term, it is envisioned to give users access to QC simulations via web-based applications, where examples of the two use cases described in the following sections will be available. 

\section{Quantum computing for classical simulation}%
\label{sec:QC4CS}


\subsection{State of the art}
\label{sec:QC4CS-SOTA}

Enhancing the simulation capabilities of scientific computing, compared to what is currently classically achievable, is one of the goals of quantum computing. This area of research can be broadly divided into two categories: quantum simulation of material and molecular properties, and QC for general linear algebra. In the QUASIM project, we focus on the latter.

Most quantum algorithms for linear algebra rely on the quantum phase estimation (QPE) subroutine \cite{kitaevbook,nielsenChuang,1998_cleve}. The problem that QPE solves can be formulated completely classically: given the eigenvector $\vec{u}$ with eigenvalue $e^{i \varphi}$ of a $N \times N$ unitary matrix $U$, estimate the phase $\varphi$ of the eigenvalue. It was shown in \cite{1999_abrams} how QPE can be turned into an eigenvalue and eigenvector finder for a general matrix $H$ with an exponential reduction in the number of systems needed, given some assumptions on the matrix $H$. We note that QPE is the basic subroutine used by Shor's factoring algorithm \cite{shor1997} and by the standard version of the Harrow-Hassidim-Lloyd (HHL) algorithm for solving linear systems of equations \cite{2009_harrow}. 

The core idea of quantum algorithms for linear algebra is to encode a $N \times N$ matrix as an operator in a quantum system involving only $\mathcal{O}(\mathrm{poly}(\log N))$ qubits. Despite this exponential compression, a quantum computer is still capable of extracting useful information about the matrix and, thus, the problem we want to solve. However, an end-to-end analysis of a quantum algorithm requires studying the scaling also with respect to other parameters such as the sparsity or the rank of the matrix, its condition number, and the desired precision of the solution. Depending on the problem, these parameters can also implicitly depend on the size of the matrix. Additionally, the encoding step \emph{in} the quantum computer and the extraction of the information \emph{from} the quantum computer, which are problem-dependent tasks, also need to be efficient. For a comprehensive overview of quantum algorithms and their scaling, please refer to \cite{dalzell2023}.

An important subroutine that is employed in many quantum algorithms is Hamiltonian simulation. In Hamiltonian simulation, we are given a Hermitian matrix $H$, which we call a Hamiltonian, and a quantum system in a state $\ket{\psi}$. The task is to transform the state into a new state evolved for a time $t$, $\ket{\psi(t)} = e^{-i H t} \ket{\psi}$, within a certain accuracy. The first quantum algorithm for Hamiltonian simulation was given by Lloyd for the case in which the Hamiltonian has an efficient decomposition in terms of operators acting locally on qubits \cite{lloyd1996}. Later, Aharonov and Ta-Shma extended this result to sparse Hamiltonians with oracle access to their matrix elements \cite{aharonov2003}. For this access model, improvements in terms of runtime and number of ancilla systems needed have been achieved during the years \cite{2007_Berry,berry2012,Childs2010}. In recent years, the seminal works of Low and Chuang \cite{lowChuang2017,lowChuang2019} allowed the formulation of the previous algorithms under the general framework of block-encoding of matrices into unitaries. The block-encoded formulation of Hamiltonian simulation further sparked the development of the quantum singular value transform (QSVT) \cite{gilyen2019,tang2023}, which, in the modern theory of quantum algorithms, is viewed as a fundamental primer of several quantum algorithms with a provable quantum advantage \cite{martyn2021}. \cite{Kikuchi2023} even provided a small scale test of Hamiltonian simulation based on block encoding on NISQ devices. On the other hand, the QSVT formulation of quantum algorithms has also allowed the inverse process of \emph{dequantization}, i.e., the development of classical algorithms inspired by the quantum ones \cite{tang2018,chia2020,gilyenLloyd2018,tang2021,bakshi2023,gharibian2022}. 

In our project, we explore the application of quantum algorithms to problems that emerge in numerical simulations of manufacturing processes. A typical example of such a problem is the analysis of the vibrations of a solid when subject to an external force. The problem translates into an eigenvalue problem that is discretized using finite element methods (FEM), for instance. Notice that once discretized the problem becomes mathematically equivalent to that of coupled harmonic oscillators, which has been recently studied in \cite{babbush2023}.  Quantum algorithms for FEM simulations have already been studied in \cite{clader2013,montanaroPallister2016}, where the authors analyzed algorithms based on the HHL algorithm \cite{2009_harrow}. General quantum algorithms for partial differential equations have also been recently investigated \cite{Berry2017,Childs2020,Childs2021highprecision,liu2021}. We remark that QPE-based algorithms are believed to require a fault-tolerant quantum computer given their high circuit depth, and as such, they are not considered NISQ algorithms. 


\subsection{General approach in QUASIM}
\label{sec:QC4CS-GeneralApproach}

State-of-the-art manufacturing processes can be optimized with the support of computer simulations, which are time-consuming and expensive.
In QUASIM, we are concerned with mathematical problems that occur frequently in industrial simulations, and quantum algorithms that solve those exact problems either in a shorter time or using less storage than any classical algorithm.
We aim for improvements on the scaling with problem-dependent parameters, such as the number of equations to solve, or the desired accuracy.
Hence, we consider computations of large mathematical problem size, which are common in simulations.

One common simulation algorithm used in manufacturing is the FEM.
This method allows the solving of physical equations for continuous systems in return for extensive computational effort.
As part of FEM, the physical domain is discretized in so-called elements.
The interaction between those elements is described by a system of numerical equations, which often require iterative solvers.
For linear equations, both, the computation time ($\mathcal{O}(N^3)$) and the required storage ($\mathcal{O}(N^2)$ ) scales polynomially with the number of equations $N$ \cite{krishnamoorthy_matrix_2013}.
For more details on the finite element method, please refer to \cite{reddy2005,huebner1975}.

In order to achieve an advantage, we concentrate on problems that can be solved with exponential speedup over their classical counterparts.
Two quantum algorithms promising such a speedup are the HHL algorithm and the QPE, which solve linear equations and eigenvalue problems, respectively.
Both problem types can be found in finite element simulations used in manufacturing.
Linear equation solvers are more versatile, which makes an end-to-end implementation of the HHL algorithm more favorable.
However, since QPE is a significant subroutine of the HHL algorithm, every knowledge gained in an end-to-end implementation of the QPE can be used for the HHL algorithm as well.
This is why we concentrate our work on the QPE and eigenvalue problems in QUASIM.

When using QPE, one has to be cautious not to lose any speedup in the subroutines such as the Hamiltonian simulation, state preparation, or post-processing steps.
Those subroutines are the interface between classical and quantum computations and scale with the amount of information that needs to enter and exit the quantum computer.
Hence, the amount of data that shall be transferred between those two platforms has to be minimal in order to contain quantum speedup ($\lesssim\mathcal{O}(N)$ for polynomial or $\lesssim\mathcal{O}(\log N)$ for exponential speed up).
Furthermore, the QPE is a probabilistic algorithm, which means that some results can only be extracted by sampling multiple times from the algorithm.

A standard application for an eigenvalue solver is the determination of eigenfrequencies of oscillating systems.
This can be of importance in various areas, such as the design of LC circuits or the prevention of manufacturing dependent excitations of vibrational eigenmodes in products \cite{ganser2021,rudel2022}.
The latter use case is described in more detail in \ref{sec:QC4CS-SpecificApproach} for the optimization of the milling process of a compressor blade.
A more detailed description of how to use QPE for the computation of vibrational eigenmodes and the corresponding response function has been developed in this project \cite{danz2024calculating}.
This work contains a full complexity analysis with respect to error tolerances and the matrix size $N$, and it explores an additional application for this algorithm.

\subsection{Specific application: milling dynamics simulation}
\label{sec:QC4CS-SpecificApproach}

Milling is a metal-cutting manufacturing process with a circular cutting motion of a usually multi-toothed tool to produce any workpiece surface.
In all milling processes, in contrast to other processes such as turning, drilling, etc., the cutting edges are not constantly engaged, but at least one cutting interruption per cutting edge occurs with each revolution of the tool \cite{klocke2018f}.  
Due to the constant cutting interruptions, depending on the milling cutter speed, a dynamic excitation of the workpiece and the tool can occur, which can have a negative effect on the surface quality in the form of vibration marks. 
Dynamic process stability simulations are carried out to analyze the vibrations and to improve the process design for the milling of thin-walled components. 
\begin{figure}[t]
    \includegraphics[trim={0 5mm 0 0 }, clip,width=1\linewidth]{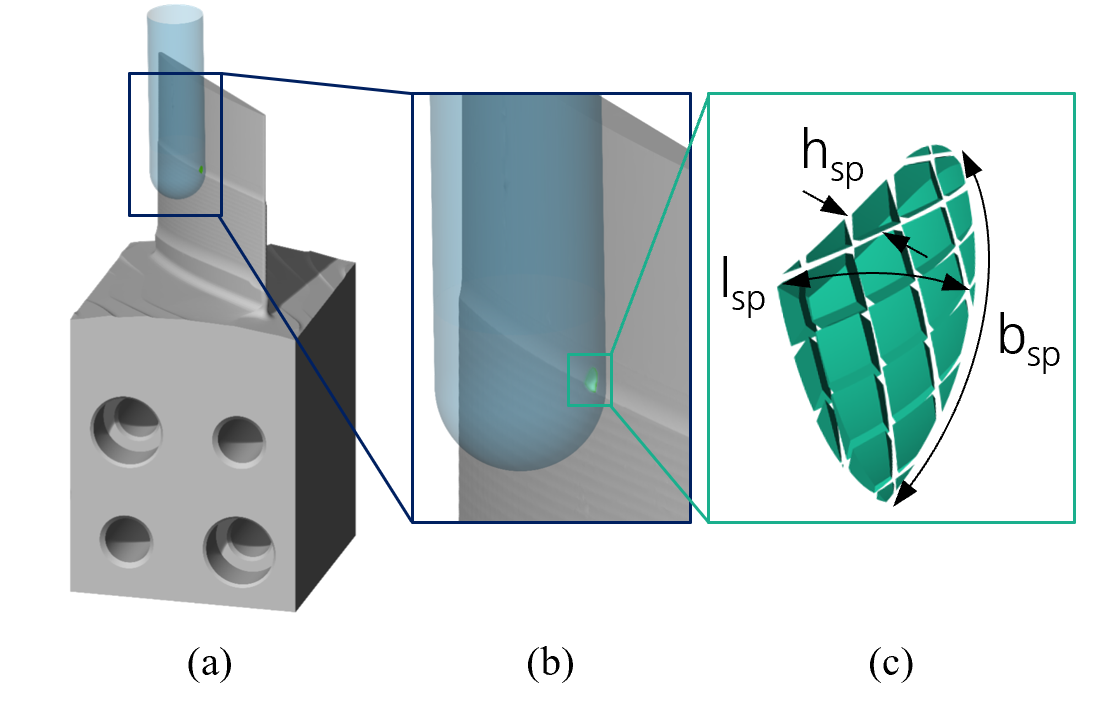}
    \caption{Simulated milling tool engagement. Shown are an in-process workpiece (IPW) of a single blade demonstrator (left), and the detailed representation of the chip removal (center) with the three characteristic chip parameters: length $l_\text{sp}$, thickness $h_\text{sp}$, and width $b_\text{sp}$.}
    \label{fig:milling}
\end{figure}

In industrial applications, models and simulations based on digital twins are mostly excluded due to their computational requirements and the expert knowledge needed to operate them. 
In industrial practice, many relevant physical effects are thus either neglected or only approximated by rough estimates. 
As a result, the quality of the digital twin, if used at all, and the insights and decisions derived from it suffer significantly, which often leads to significant economic disadvantages in the industry. 
Due to high-quality requirements and usually considerable costs for rejects, simulations based on digital twins enable economically and ecologically optimized machining processes to be planned and executed.


Within the digital twin framework of the Fraunhofer Institute for Productions Technology Aachen dPart\textsuperscript{\textregistered} \cite{ganser2021}, the engagement simulation is a dexel-based simulation. 
A dexel model is a discrete representation of a solid volume based on a regular grid, which enables a more efficient simulation of material removal processes than mesh-based methods.
For a good representation of 3D volumes, a multi-dexel approximation is used \cite{stautner2006simulation}.
The engagement simulation provides the tool's contact area, contact length, and width (see Fig.~\ref{fig:milling}). 
It also allows tracking removed volume to enable an evaluation of the workpiece cut by cut.
To compute the removed material, the algorithm calculates the intersections between the milling tool volume and the three-dimensional dexel field.
The intersections are performed individually for each of the three dexel grids to take advantage of memory closeness. 
The calculations performed by the intersectors are based on floating-point calculations.

Modeling mechanical and thermal effects during machining, such as the calculation of dominant frequencies of thin-walled workpieces based on process-specific boundary conditions, is frequently accomplished via FEM \cite{19.Scippa_A,20.Wang_D,21.Budak_E}. The calculation of tool and workpiece dynamics at each selected cutter location along the toolpath is essential to describe vibrations occurring during the cutting process \cite{22._Wang_X,23._Bachrathy_D,24._Biermann_D}. 
Within novel simulation approaches based on FEM and computer-aided manufacturing (CAM), it is possible to take into account systematically the continuously changing workpiece dynamics due to material removal (see Fig.~\ref{fig:milling}) \cite{25_Maslo_S,26_Maslo_S}. 
This allows a position-oriented evaluation of the frequency response function (FRF) of the workpiece, which can be subsequently coupled with simulated or experimentally determined milling tool FRFs measured at the tooltip. 
The FEM simulation workflow, originally published in \cite{rudel2022}, is performed for each in-process workpiece (IPW) (Fig.~\ref{fig:milling}) in subsequent calculations. 

Starting after the dexel-based engagement simulation in CAM, the main steps of the FEM simulation workflow are: IPW conversion from dexel to solid, meshing process, and modal analysis. 
Considering the condition for quantum acceleration, the workflow is analyzed step by step both in terms of computational time and with regard to extracting numerical problems that can be transformed in quantum algorithms \cite{schroeder24}.  
The “modal analysis”, i.e., solving the eigenvalue problem, is found to be a suitable numerical problem for computation on a quantum computer. 
The solution of the eigenvalue problem using quantum algorithms is under investigation within this project.
Within the classical step ``modal analysis" the equation of motion 
\begin{equation}
     M  \ddot{x}(t) +  K  x(t)\ = {0},
\end{equation}
for a free and undamped system is consulted in the physical domain on a local simulation computer. 
Here, $M$ is the mass matrix, $K$ is the stiffness matrix and $x(t)$ and $\ddot{x}(t)$ contain the displacements.
In combination with the equations of harmonic motion, the equation of motion can be transformed into a generalized eigenvalue problem
\begin{equation}
    (K - \omega_i^2 M) \Phi_i = {0}.
\end{equation}


The eigenvalues and eigenvectors are the square of natural frequency $\omega_i^2$ and the mode shapes $\Phi_i$ of the structure, respectively. By knowing eigenvalues and eigenvectors (and damping ratio), the dynamics for each simulated IPW can be described in state space representation, usually visualized as FRF on the quantum computer. 
The calculated solutions can then be sent back to the local simulation computer via a quantum as a service backend. 
There, the results can be further utilized for optimization of the spindle speed to reduce process vibrations.

\section{Quantum machine learning for classical simulation}
\label{sec:QML}


\subsection{State of the art}
\label{sec:QML-SOTA}

In recent years, ML has made significant progress due to a combination of better hardware as well as new processing techniques, both of which have allowed models to process larger amounts of data than ever before \cite{vaswani2017attention,krizhevsky2012imagenet,goodfellow2016deep}. 
In addition to typical data-driven scenarios, ML has also become an alternative to numerical solutions for the simulations of classical physics \cite{karniadakis2021physics}. 
A major challenge of state-of-the-art ML is the increasing amount of computational resources needed for training. 
One promising way to mitigate this challenge is to use quantum computing for more efficient learning methods.

Since NISQ computers are still in their infancy, and limited in size and available gate depths, they often can be simulated on classical hardware.
Therefore, one can use real and simulated NISQ computers alongside classical computers to perform tests on machine learning tasks in hybrid QC. 
A popular implementation of this uses variational or parameterized quantum circuits (PQCs), which have an advantage over classical ML in certain learning tasks \cite{bowles2023contextuality}.

Such PQCs generally consist of four parts: encoding blocks, parameterized blocks, a measurement, and a loss function. The encoding blocks map classical data into a quantum state, while parameterized blocks consist of quantum gates with tunable parameters that are optimized throughout the learning process. The choice and structure of gates has a large effect on the outcome of the learning process \cite{perez2020data,havlivcek2019supervised,schuld2021effect,skolik_equivariant_2023}. The measurement collapses the qubits into one of the computational basis states and passes an average over several measurement outcomes to a classical computer, where the loss function is calculated, and the optimization step is performed. The parameters in the circuit are then tuned to minimize the loss function \cite{schuld2015,benedetti2019parameterized}. 

Below, we demonstrate how large graph data sets can be processed using PQCs. Section \ref{sec:QML-SpecificApproach} then demonstrates an application of these methods to the laser cutting use case.

\subsection{General approach in QUASIM}
\label{sec:QML-GeneralApproach}

In the simulation of manufacturing processes, meshes are used to digitally represent a space-discretized version of produced parts.
We focus on simulated physical quantities stored in a mesh-based data structure.
One can view the meshes as graphs $G=(V, E)$, with $V$ the set of vertices, $E$ the set of edges, and represent the physical quantities of the manufactured part as node features $f_v^t$ and edge features $\varepsilon_{v,w}^t$, for vertices $v,w$, and for a certain timestamp $t$. The task is to predict an update for the physical quantity of interest, represented by the label $f_v^{t+1}$, for each vertex of the graph (for simplicity, we focus on vertex-based data structures).

For such data structures, graph neural networks (GNNs) are the preferred machine learning approach, since they can leverage the symmetries of the graph. 
There are quantum variants of such approaches that take symmetries into account, for example, equivariant models for unweighted \cite{mernyei2022equivariant} and weighted graphs \cite{skolik_equivariant_2023}. Equivariant models permute the outputs in the same way as the inputs are permuted.
The downside of these proposals is that at least one qubit is needed for each node and feature, thus rendering large graph data sets unfeasible for near-term quantum computers. 

In order to make such methods amenable to NISQ, we turn to a local model:
instead of passing the whole graph to a PQC the circuit is parsed centered at one vertex at a time, which effectively lowers the input size for the circuit. 
In this approach, an update consists of applying many submodels $M_d$ in parallel, each to a tree subgraph $G_v$ of $G$. The set of vertices of $G_v$ consists of $v$ as the root node and its neighbors, $\mathcal{N}(v)$, as leafs, having the root vertex degree $d=|\mathcal{N}(v)|$. Submodels (together with their weights $\vec{\theta}_d$) are shared across the subgraphs with the same degree.
We restrict ourselves to nearest neighbors here, which turns the tree into a star graph. The extension beyond the nearest neighbors is straightforward. 


Every submodel creates a quantum state $\ket{v,\mathcal{N}(v), \vec{\theta}_d}$, via the application of gates parameterized by learnable parameters and data features. 
The expectation values of the quantum state with respect to an observable $O$ are obtained through measurement and decoded using a classical function $D$. 
For the label $f_v^{t+1}$ at node $v$, the prediction $\tilde{f}_v^{t+1}$ is obtained via the equation

\begin{equation}\label{eqn:qgnn_prediction}
\begin{split}
        \tilde{f}_v^{t+1} &= M_d (\vec{\theta}_d, f_v^t, f_{w_1}^t, \dots, f_{w_d}^t, \varepsilon_{v,w_1}^t,\dots,\varepsilon_{v, w_d}^t)\\
    &=D \left( \bra{v,\mathcal{N}(v), \vec{\theta}_d}O\ket{v,\mathcal{N}(v), \vec{\theta}_d} \right). 
\end{split}
\end{equation}

The specific choices of $D, O$, and the circuit ansatz used to construct the state $\ket{v,\mathcal{N}(v), \vec{\theta}_d}$ can be problem-dependent and need to be specified accordingly. However, we work with a few guidelines for defining these local models: parameterized one-qubit gates can be used for the encoding of node features, while two-qubit gates can entangle qubits that refer to connected nodes, parameterized by learnable parameters and edge features.

\begin{figure*}[t]
\centering
    \includegraphics[width=0.8\linewidth]{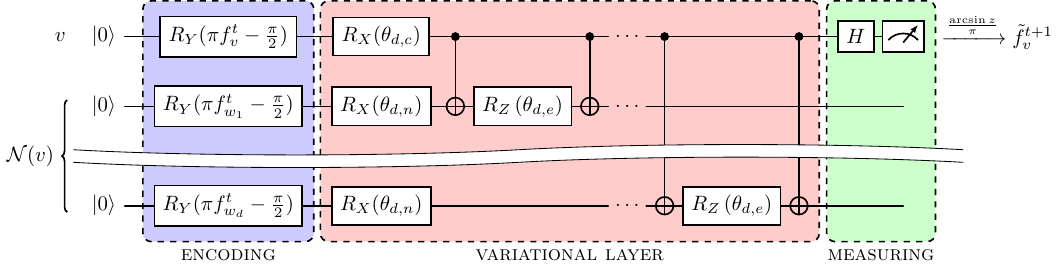}
     \caption{The PQC $M_d$ used to make temperature predictions in the laser cutting use case. The first qubit represents $v$, and the subsequent qubits represent the $d$ neighboring nodes. The temperature of each node is embedded on to the corresponding qubit. This is followed by a parameterized layer, where the first qubit and the subsequent qubits are entangled. Finally, the expectation value of the first qubit is measured and then projected to a temperature value.}
    \label{fig:QML_example_circuit}
\end{figure*}

The submodels must be made invariant under permutations of the neighbors so that the order in which the nodes are read does not affect the output of the submodel. This reflects one symmetry of scalar functions defined on graphs. This inductive bias can be ensured by choosing the appropriate gate architecture and reusing parameters. Figure~\ref{fig:QML_example_circuit} shows an example, explained in the next section.


Note that the combination of locality and degree-focus turns a trained model reusable for parts with different geometry, as long as the range of degrees of their meshes are equal.

\subsection{Specific application: laser cutting}
\label{sec:QML-SpecificApproach}

This section delves into the promising potential of PQCs for addressing real-world manufacturing challenges currently addressed by FEM simulations. The example of heat distribution management in laser cutting processes with a TRUMPF laser cutting system serves to illustrate this potential. 

TRUMPF laser cutting machines possess the capability to operate autonomously once the desired cutting program is specified, handling everything from running the laser path to loading or unloading and sorting materials. This enables the machine to operate without human supervision, yet frequent production standstills persist due to thermal expansion on laser-cut parts, causing up to 1\% of parts getting stuck inside the sheet and having to be removed manually. Consequently, laser cutting machines can run into a standstill overnight, which results in significant economic disadvantage. To solve this problem, it is proposed to simulate the heat distribution during the laser-cutting process. By accurately modeling it and subsequently predicting thermal expansion, an optimized cutting route that minimizes the risk of machine standstills can be planned.  


Accurately simulating the temperature during laser cutting with complex geometries and long cutting times requires significant computational efforts. In contrast to conventional FEM-based simulations, ML approaches, such as MeshGraphNets \cite{pfaff2020mgn}, accelerate FEM simulations by one to two orders of magnitude whilst maintaining a high accuracy. However, training such large ML models currently takes several weeks, which makes updating such models computationally challenging. 
Efforts have been made to conduct thermal simulations using hybrid-QML approaches, such as QFE-GNNs \cite{XuSascha2024QuantumFE}, where the encoding layer of the classical GNN is substituted with PQC to enhance feature representation. This method achieves better performance than traditional GNNs, however, scaling these models to handle larger graphs with complex geometries remains challenging.
In this work, the advantages of quantum machine learning approaches, as described above, are investigated to simulate thermodynamics from laser-induced heat on metal plates.



The metal geometry is represented by a mesh structure, which can be formulated as a graph $G=(V, E)$. The model is formulated employing the PQC approach outlined in section \ref{sec:QML-GeneralApproach} for predicting the temperature $f_v^{t+1}$ at the vertex $v$ for the next time step $(t+1)$, depending on the temperatures from the previous time step $t$ at the vertex $v$ and its $d$ neighbors $\mathcal{N}(v)$. A visualization of the circuit can be seen in Fig.~\ref{fig:QML_example_circuit}.



Reusing the parameters $\theta_{d,n}$ and $\theta_{d,e}$ across wires representing neighbors turns the circuit invariant under permutations of the neighbors, in the sense that the same temperature prediction $\tilde{f}_v^{t+1}$ is obtained for any order in which the relative temperatures $f^t_{w_k}$ are fed into qubits.



For a fixed graph $G$ and time step $t$, the mean squared error (MSE) is minimized, which can be computed as
\begin{equation}
\ell_{G, t}(\theta) = \frac{1}{|V|}\sum_{v\in V} \left|f^{t+1}_v - \left[M_\theta(G, \{f^{t}_v\})\right]_v\right|^2,
\end{equation}
where, $\theta$ denotes the trainable parameters and $M_\theta$ represents our overall model.

Future enhancements to the model will focus on integrating additional features, such as the distance between nodes, the thermal properties of the metal being cut, and the trajectory of the laser. There are also plans to investigate different circuit ansatzes, including methods for data re-uploading and a variety of circuit architectures, to produce more accurate predictions.



\section{Summary and Outlook}
\label{sec:outlook}


The quantum computing project QUASIM represents a pioneering effort to integrate QC with traditional manufacturing processes, focusing on milling and laser-cutting use cases. This endeavor marks a significant advancement in the application of quantum technology in high-value manufacturing operations. The project has addressed these complex processes by combining domain-specific knowledge with QC expertise, offering new insights into potential efficiencies and precision enhancements. This synergy of traditional manufacturing techniques and cutting-edge quantum computation is a testament to the project's innovative approach and potential to revolutionize industry practices.

In the short term, heuristic benchmarking to analyze the expected quantum advantages in manufacturing simulations needs to be targeted. This involves a detailed investigation into the resource requirements of NISQ technologies \cite{Preskill18}, focusing on evaluating the performance of graph neural networks against phase estimation methods in quantum contexts. These initial steps are crucial in establishing a solid foundation for QC applications in manufacturing, ensuring the technology is not only advanced but also relevant and applicable to industry needs.


Looking ahead, the project envisions the development of comprehensive, end-to-end software packages that seamlessly integrate quantum algorithms with manufacturing processes. These software solutions aim to be user-friendly and accessible to industry professionals, bridging the gap between QC theory and practical application. In the longer term, the creation of application prototypes is planned. These prototypes will serve as tangible examples of quantum computing's practical application in real-world manufacturing scenarios, demonstrating its effectiveness and paving the way for broader adoption in the industry. Through these ambitious goals, the project aims to lead the way in applying quantum computing to enhance high-tech manufacturing processes, setting a precedent for future innovations in the field.


%
%

\bibliographystyle{spmpsci}      

\bibliography{bibliography}

\end{document}